\documentstyle[prl,multicol,epsf,aps]{revtex}
\draft

\title{Anomalous  diffusion and Tsallis statistics in an optical lattice}
\author{Eric Lutz}
\address{Sloane Physics Laboratory, Yale University, P.O. Box 208120, New Haven,
CT 06520-8120, USA}

\date{\today}

\def\openone{\leavevmode\hbox{\small1\kern-3.3pt\normalsize1}}
\newcommand{\la}{\langle}
\newcommand{\ra}{\rangle}
\newcommand{\be}{\begin{equation}}
\newcommand{\ee}{\end{equation}}
\newcommand{\bea}{\begin{eqnarray}}
\newcommand{\eea}{\end{eqnarray}}
\newcommand{\pd}{\partial}
\newcommand{\SC}{\scriptstyle}

\begin{document}

\twocolumn[\hsize\textwidth\columnwidth\hsize\csname@twocolumnfalse\endcsname

\maketitle \vspace{-0.2cm}
\begin{abstract}
We point out a connection between anomalous quantum transport in an optical lattice and
 Tsallis' generalized thermostatistics.
Specifically, we show that the momentum equation for  the
semiclassical Wigner function that describes atomic motion in the
optical potential, belongs to a class of transport equations
recently studied by Borland  $\mbox{[}$PLA  245, 67
(1998)$\mbox{]}$. The important property of these ordinary linear
Fokker--Planck equations is that their stationary solutions are
exactly  given by  Tsallis distributions. Dissipative optical
lattices are therefore new systems in which Tsallis statistics can
be experimentally studied.
\end{abstract}
\pacs{PACS numbers: 05.60.-k, 42.50.Vk} \vskip.5pc]

 Non-Gaussian
distributions occur frequently in systems that do not follow the
prescriptions of standard statistical mechanics. Prominent
examples of non--Gaussian statistics  are  L\'evy stable
distributions  \cite{shl93}. The probability density of a
one--dimensional symmetric L\'evy stable  distribution is defined
by its Fourier transform as ${\cal
L}^{\SC{C}}_\alpha(x)\!=\!1/(2\pi) \int dk \,\exp[ikx-C
|k|^\alpha]$, where $0\!<\!\alpha\! \leq \!2$. A key feature of
such a stable distribution is the presence of an asymptotic,
non--Gaussian,  power--law tail, ${\cal L}^{\SC{C}}_\alpha(x) \sim
1/|x|^{\alpha+1}$,  when $\alpha\!<\!2$. This leads to the
important consequence that, except in the Gaussian case
$\alpha\!=\!2$, a L\'evy probability density has a divergent
second moment \cite{lev37}. Signatures of L\'evy statistics have
been experimentally observed in a variety of physical systems
\cite{shl94} ranging from micelle systems \cite{ott90} to porous
glasses \cite{sta95}  and subrecoil laser cooling \cite{bar02}.
Another non--Gaussian distribution, which naturally arises within
the framework of nonextensive statistical mechanics
\cite{tsa88,abe01}, is the Tsallis
 distribution,
$P_q(x) = Z_q^{-1} [1-\beta(1-q)x^2]^{1/(1-q)}$ with  $1\!\leq
\!q\!<\!3$ \cite{rem}. Like a L\'evy stable distribution, the
function $P_q(x)$ exhibits an asymptotic, non--Gaussian, algebraic
tail, $P_q(x) \sim 1/ x^{2/(q-1)}$, for a Tsallis index $q\!\neq
\!1$. Typical systems where Tsallis' generalized thermostatistics
has been applied  are those involving long--range correlations,
such as self--gravitating systems \cite{abe01} or long--range
magnetic systems \cite{rei02}, and systems with fluctuating
temperature \cite{bec02}. In the last decade the theory  of
nonextensive statistical physics witnessed a tremendous
development \cite{tsa} and there is now also growing experimental
evidence of the relevance of Tsallis statistics in  describing
physical processes \cite{abe01}. As an example we mention fully
developed turbulence, where it has recently been shown that
velocity fluctuations are well described by a Tsallis--like
distribution \cite{bec01}. However, there is a  need for more
experimental support.

Our aim in this paper is to show that there is a connection
between Tsallis  statistics and anomalous quantum transport in
optical lattices. An optical lattice is a standing--wave potential
that can be  obtained by superposition of counter--propagating
laser beams with linear orthogonal polarizations (for a recent
review see \cite{gry01}). The  optical  potential so produced is
spatially periodic and, as a consequence,  shares many common
properties with crystalline  lattices in solid--state physics,
such as  Bragg scattering and Bloch oscillations. The main
advantage of an optical crystal compared to its condensed--matter
counterpart is that the optical periodic potential is exactly
known and, furthermore, easily modified in a precise and
controlled way. Originally designed for laser cooling (Sisyphus
cooling)(for an introduction see \cite{met99}), optical lattices
rapidly evolved to an active field of investigation of its own
\cite{gry01}.

An important issue in this context is the  understanding of atomic
transport in the optical lattice. Depending on the depth of the
optical potential, three different regimes can be identified
\cite{gry01,cas91,hod95,jur96,mar96,kat97}: (i)  diffusive motion
in deep potentials, (ii) ballistic motion in shallow potentials
and (iii)  an intermediate regime in between (of main interest
here)  where anomalous (non--Gaussian) diffusion takes place. The
existence of L\'evy--like diffusion with long jumps below a given
potential threshold has been predicted by Marksteiner {\it et al.}
\cite{mar96} and later experimentally verified  by a  group at the
MPQ in Garching by  studying the dynamics of a single ion in a
one--dimensional optical lattice \cite{kat97}. In the following,
we show that the  equation governing the evolution of the
semiclassical momentum distribution of the atom in the optical
potential belongs to a family of ordinary linear Fokker--Planck
equations recently defined by Borland \cite{bor98}. The
interesting property of these equations is that their stationary
solutions are  exactly given by Tsallis statistics. This allows us
not only to express the indices $q$ and $\beta$ of the Tsallis
distribution in terms of the microscopic parameters of the quantum
optical problem, but also to give a physical explanation for the
non-normalizability of the distribution, as well as for the
divergence of its  variance in some range of parameters to be
specified. We also introduce the Rayleigh equation for a  quantum
L\'evy process and compare the properties of its stationary
solution with those of the Tsallis distribution. We find that the
two processes lead to different, experimentally observable,
predictions.

Starting from the microscopic Hamiltonian that describes the
atom--laser interaction in the optical lattice, an atomic quantum
master equation can be derived \cite{dal89}. After spatial
averaging, the Rayleigh equation for the corresponding
semiclassical  Wigner function $W(p,t)$  can be written as
\cite{cas91,hod95,mar96},
\be
\label{eq1}
\frac{\pd W}{\pd t} = -\frac{\pd}{\pd p}\left[K(p)W\right] + \frac{\pd }{\pd p}\left[D(p)\frac{\pd W}{\pd p}\right] \ .
\ee
Equation (\ref{eq1}) has the form of an  ordinary linear Fokker--Planck equation with momentum--dependent drift and diffusion coefficients,
\be
\label{eq2}
K(p) = -\frac{\alpha p}{1+(p/p_c)^2}, \hspace{0.3cm} D(p) = D_0 + \frac{D_1}{1+(p/p_c)^2} \ .
\ee
These two quantities have a simple physical interpretation:
The drift $K(p)$ represents a cooling force (due to the Sisyphus effect) with damping coefficient  $\alpha$. This force acts only on slow particles with a momentum smaller than the capture momentum  $p_c$. This is an important point as we shall discuss below. The diffusion factor $D(p)$, on the other hand, describes stochastic momentum fluctuations and accounts for heating processes.  We note that $D(p)$ has two contributions \cite{cas91}: A  constant part $D_0$ which corresponds to fluctuations due to spontaneous photon emissions and fluctuations in the difference of photons absorbed in the two laser beams,  plus a term proportional to $D_1$ which stems from  fluctuations in the dipolar forces. This last term has the same limited momentum range  $p_c$ as the drift force. Interestingly, we remark that  for  vanishing  $D_1$, Eq.~(\ref{eq1}) exactly reduces to the Fokker--Planck equation  considered by Stariolo and shown to give rise to nonextensive statistics \cite{sta94}.

\begin{figure}[t]
\centerline{\epsfxsize=7.0cm \epsfbox{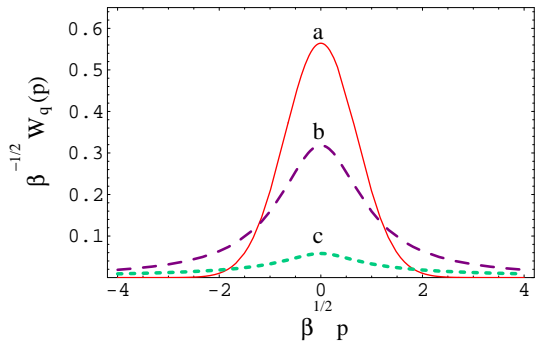}} \vspace{.1cm}

\caption{Tsallis distribution $W_q(p)$ (\ref{eq4}) for three  values of the optical potential: (a) $U_0 \gg 44 E_R$, (b) $U_0=44 E_R$ and (c) $U_0 = 24 E_R$.}
\label{fig1}
\end{figure}

It is  easily seen from Eq.~(\ref{eq2}) that $K$ and $D$ satisfy the following
condition,
\be
\label{eq3}
\frac{K(p)}{D(p)} = -\frac{\beta}{1-\beta(1-q) U(p) }\frac{\pd U(p)}{\pd p} \ ,
\ee
with
\be
\label{eq31}
 \beta= \frac{\alpha}{2(D_0+D_1)}\ ,\:\: q=1+\frac{2D_0}{\alpha p_c^2}\: \mbox{ and }
\:U(p) = p^2 \ .
\ee
Equation (\ref{eq3}) has been first obtained by Borland \cite{bor98}.
We mention that in her original work, Borland considered the Ito--form of the Fokker--Planck equation, whereas here, Eq.~(\ref{eq3}) applies to the Stratonovich--form (\ref{eq1}). The condition (\ref{eq3}) implies that the stationary solution $W_q(p)$ of the Rayleigh equation (\ref{eq1})
 is given by the Tsallis distribution,
\be
\label{eq4}
W_q(p) = Z_q^{-1}[1-\beta(1-q)U(p)]^{1/(1-q)}\ .
\ee
Equation (\ref{eq4}) is the exact  general stationary solution of
Eq.~(\ref{eq1}) with the requirement $W_q(p) \rightarrow 0$ when
$p\rightarrow \pm \infty$, the constant $Z_q$ being a normalizing
factor. The fact that the steady--state solution of
Eq.~(\ref{eq1}) is non--Gaussian is of course well-known
\cite{cas91,hod95,mar96}. Surprisingly, however, it has not been
realized  that this {\it precisely} corresponds to a Tsallis
distribution. Among infinitely many non--Gaussian distributions,
Eq.~(\ref{eq3}) singles out the nonextensive Tsallis distribution
(\ref{eq4}) (see Fig.~\ref{fig1}). It is worth noting that  the
Tsallis indices $q$ and $\beta$ can be simply expressed in terms
of  the microscopic parameters of the problem [see
Eqs.~(\ref{eq31})]. In particular, we see that $q$ depends on the
ratio of the diffusion constant $D_0$ to the product  of the
friction coefficient $\alpha$ with the square of  the capture
momentum $p_c$, and does not depend on $D_1$. Equations
(\ref{eq31}) thus provide  a link between the macroscopic Tsallis
distribution (\ref{eq4}) and the underlying microscopic dynamics
in the optical potential. This allows us to give a physical
interpretation of the characteristics of the nonextensive
distribution (\ref{eq4}).

Let us first  remind that  the distribution (\ref{eq4}) is not
normalizable for a Tsallis index  $3 \!\leq \!q$ or, equivalently,
for $\alpha p_c^2\!\leq \!D_0$. Physically, this means that the
cooling force is too weak, compared to the random momentum
fluctuations, to maintain the particle in a steady state around
$p=0$ (this is often referred to as {\it d\'ecrochage}
\cite{cas91,hod95}). On the other hand,  in the limit where
$q\!\rightarrow\!1$ ($D_0\!\ll\!\alpha p_c^2$), the stationary
solution (\ref{eq4}) reduces to the standard Maxwell--Boltzmann
distribution, $W_1(p)= Z_1^{-1} \exp[-\beta U(p)]$,  with an
inverse temperature $\beta$. In this case the cooling force is
much stronger than the random momentum fluctuations. It thus
appears that the Tsallis index $q$ is intimately related to the
interplay between stochastic heating processes (momentum
fluctuations, as measured by $D_0$) and  the cooling force with
capture momentum $p_c$. It is important to remark that the
finiteness of the latter is directly responsible for the
occurrence of the non--Gaussian Tsallis distribution in this
problem. This is confirmed by the observation that for infinite
$p_c$, Eq.~(\ref{eq1}) reduces to the Ornstein--Uhlenbeck equation
with  well--known Gaussian dynamics. Using the parametrization of
Ref.~\cite{cas91}, the index $q$ can be further written as $q=1+44
E_R/U_0$, where $U_0$ is the potential depth and $E_R$ the recoil
energy. We  thus see that the Tsallis index can be related to  the
ratio of the recoil energy to the potential depth. This means that
the nature of the atomic dynamics can be simply tuned by varying
the depth of the optical lattice. We also notice  that the inverse
temperature $\beta$ is written as the ratio of the friction
coefficient to the sum of the diffusion coefficients, in analogy
with the fluctuation--dissipation relation. We hasten to add that
Eq.~(\ref{eq4}) corresponds to a steady state and  not to an
equilibrium state, and as such temperature is not well--defined in
this problem.

We now turn to the intermediate regime with a  Tsallis index  $5/3\!<\!q\!<\!3$ ($D_0\!<\!\alpha p_c^2\!<\!3D_0)$. Here  the second moment, $\la p^2 \ra = \int p^2 W_q(p) dp$, of the Tsallis distribution is infinite.  As a consequence, the mean kinetic energy of the particle,  $E_K= \la p^2\ra/2m$, diverges. In this regime, rare but large  momentum fluctuations occur that shove the particle outside the range of the cooling force before it is  recaptured again. This leads to anomalous momentum diffusion. The transition from Gaussian to anomalous diffusion as the depth of the optical  lattice  is decreased has  recently been investigated experimentally  and the divergence of the mean kinetic energy has been observed  \cite{kat97}. This is a clear signature of the underlying  non--Gaussian statistics. A dissipative optical lattice hence  appears as a unique system that allows  experimental investigation of  the Tsallis distribution in a whole range of  $q$ by simply varying a single parameter, the  depth of the optical potential.

We emphasize that  the non--Gaussian Tsallis statistics is here generated by an {\it ordinary linear} Fokker--Planck equation (\ref{eq3}), which is often associated with the usual Boltzmann--Gibbs statistics. To our knowledge, atomic  transport in an  optical lattice constitutes the only  physical system known so far where this occurs. Again, this results from the subtle interplay between the deterministic (drift) and stochastic (diffusion) forces (\ref{eq2}) that act on the particle \cite{bor98}. This is for instance at variance with the fully developed turbulence problem discussed in  Ref.~\cite{bec01}. In the latter case,   nonextensive statistics  is    obtained from a generalized Langevin equation with {\it fluctuating} friction and diffusion coefficients. For comparison, the Langevin equation that corresponds  to the Rayleigh equation (\ref{eq1}) reads
\be
\label{eq5}
\dot p = K(p)+ \frac{\pd D(p)}{\pd p} + \sqrt{2 D(p)}\, \eta(t) \ ,
\ee
where $\eta(t)$ is a centered {\it Gaussian} random force with variance $\la\eta(t)\eta(t')\ra=\delta(t-t')$. Equation (\ref{eq5}) is a Langevin equation with multiplicative  white noise and {\it deterministic} coefficients.

\noindent

Anomalous diffusion in optical lattices shares common properties
with   a L\'evy process and is often associated with it
\cite{mar96,kat97}. So let us next compare in some detail the two
quantum processes based on Tsallis and L\'evy statistics. The
Klein--Kramers equation for a quantum L\'evy process has been
recently derived in \cite{lut01} using a path--integral
approach.  The corresponding Rayleigh equation for a free quantum
L\'evy process can  be obtained by integrating over the position,
\be
\label{eq6}
\frac{\pd W}{\pd t} = \frac{\gamma}{m} \frac{\pd}{\pd p}[p W]+ D \frac{\partial^\alpha
  W}{\partial |p|^\alpha} \ ,
\ee
where $\gamma$ and $D$ respectively denote the friction and diffusion coefficients.
In Eq.~(\ref{eq6}), we have introduced the Riesz fractional derivative which is defined
through its Fourier transform as \cite{sam93,sai97},
\be
\label{eq7}
 -\frac{\partial^\alpha}{\partial
  |p|^\alpha} = \frac{1}{2\pi}\int_{-\infty}^\infty dy\,
\exp[-ipy]\, |y|^\alpha\ .
\ee
We point out that contrary to  the ordinary Fokker--Planck equation (\ref{eq1}), Eq.~(\ref{eq6}) is a  generalized, fractional transport equation. It reduces to the Ornstein--Uhlenbeck equation when $\alpha=2$.
The stationary solution of Eq.~(\ref{eq6}) can be  easily found by Fourier  transformation  \cite{jes99}. It is given  by the L\'evy distribution,
$W_{\alpha}(p)={\cal L}^{\SC{C}}_\alpha(p)$ with $C= m D/\alpha \gamma$. As already indicated in the introduction, the function $W_\alpha(p)$ asymptotically decays according to a power--law, $W_\alpha(p)\sim 1/|p|^{\alpha+1}$. This is reminiscent of a Tsallis distribution with index $q=(3+\alpha)/(1+\alpha)$. However, there is an  important  difference between a L\'evy and a Tsallis distribution: In the L\'evy case, the second moment is finite  for only one {\it single} value of the parameter $\alpha=2$. By direct contrast, in the case of a Tsallis distribution, the second moment is finite in a  parameter  {\it interval} $1\!<q<\!5/3$.  This implies that the mean kinetic energy of the Tsallis  particle is finite over a complete range of potential depths and not only for a particular value of $U_0$, as this would be the case for a particle obeying L\'evy statistics. This is in agreement with what is measured experimentally [see Figure (4) of Ref.~\cite{kat97}].

It is further  instructive to write down the Langevin equation
that corresponds to the Rayleigh equation (\ref{eq6}). It reads
\be
\dot p = -\frac{\gamma}{m}\, p + \xi(t) \ .
\ee
Here $\xi(t)$ is a centered {\it  L\'evy} distributed stochastic
force. Its  characteristic function is given by $\tilde P(k)= \int
d\xi \exp[i k \xi] P(\xi) = \exp[-D|k|^\alpha]$. We see that for a
L\'evy process, the anomalous behavior of the particle  finds its
origin in the non--Gaussian properties of the noise. This is  in
contrast to the Langevin equation (\ref{eq5}) where the noise is
Gaussian. From all this we conclude that the strange kinetics that
occurs in an optical lattice, in the semiclassical limit
considered here, does not correspond to a quantum L\'evy process,
strictly speaking, but rather a ``quantum Tsallis process''.

Interestingly,  broad momentum distributions and L\'evy statistics
 play an important role in another branch of atomic physics,
namely subrecoil laser cooling --- L\'evy statistics being even
used as a tool to optimize the  cooling process \cite{bar02}. The
physical mechanism that leads here to power--law momentum
distributions of the cold atoms is based on the presence of a
trap, $|p|<p_{\SC{trap}}$, around $p=0$ where particles remain for
a long time during their random walk in momentum space, before
leaving again. The broad momentum distribution can then be seen as
resulting from the competition between the rates of entry and of
departure in the trap \cite{bar02}.

In conclusion, we have shown that  Tsallis  statistics  naturally
appears in anomalous quantum transport in an optical lattice.  Remarkably the  Tsallis distribution
is here generated by an ordinary linear Fokker--Planck equation
and not by some generalized (non--linear) diffusion equations.
Furthermore, the Tsallis index $q$ can be simply expressed  in
terms of the microscopic parameters of the quantum--optical
problem, in particular, the  potential depth $U_0$. This shows
that the shape of the  distribution  can be straightforwardly
modified --- from a Gaussian to a uniform distribution ---  by
solely varying  $U_0$. This opens the possibility of
experimentally studying Tsallis distributions and non--Gaussian
dynamics for any index  $q$.

We thank Lisa Borland and  Hidetoshi  Katori for discussions. This
work was funded in part  by the ONR under contract
N00014-01-1-0594.

\end{document}